# Investigation of Tunable Structured Light Using Bilayer Parity-Time Symmetry Dammann Grating Metasurfaces

Xiang Cai[1], Zhiwei Shi[1],*, Wei Liu[1], Zhen Yao[1], Huagang Li[2,†],

Yaohua Deng[1,#]

1  School of Electromechanical Engineering, Guangdong University of Technology, Guangzhou 510006, China

2  School of Electronics and Information, Guangdong Polytechnic Normal University, Guangzhou 510665, China

Correspondence: ∗ szwstar@gdut.edu.cn,

†lhg_3@sina.com , #dengyaohua@gdut.edu.cn

## Abstract

In the current technological landscape, structured light technology holds a critically important position. However, traditional structured light optical components often require complex systems and extensive resources for application, and they function in a fixed manner. This study takes this challenge as an opportunity to design a novel dynamically tunable double-layer Dammann grating (DG) metasurface. During the research, we developed a double-layer DG metasurface structure using silica as the substrate and lithium niobate (LiNbO3, LN) as the nanocolumn material. By specifically introducing parity-time (PT) symmetry, we designed three distinct states, combined with rotational transformations leveraging the Moiré effect. Further investigations revealed that for metasurfaces with different radius combinations, changes in rotation and PT symmetry states resulted in significant variations in the shape, position, and intensity of the diffraction spots, alongside changes in conversion efficiency and contrast ratio. This study thoroughly and comprehensively unveils the significant impacts of rotational transformations, PT symmetry, and radius combination on the optical characteristics of double-layer DG metasurfaces, providing a new method for the design of dynamic tunable optical components with structured light.

## 1. Introduction

Structured light imaging, as a precise method for 3D visualization, captures intricate surface details and spatial coordinates of objects, finding applications across numerous fields. These include industrial inspection [1] and robotic vision systems [2] in precision manufacturing, as well as advanced technologies such as minimally invasive surgical navigation [3] and biometric identification [4] in medical diagnostics. Additionally, structured light techniques also offer unique value in augmented reality [5], art restoration [6], and security surveillance [7]. With technological advancements, structured light imaging continues to evolve with innovations such as multi-view imaging techniques [8] and dynamic light modulation technologies [9], enhancing both the speed and flexibility of imaging and opening new possibilities for accurate 3D reconstruction in complex application environments.

Within the context of structured light technology, it is essential to mention binary optics technology, and a quintessentially important element within this is the DG. Originating in the 1970s, the DG was proposed by the scientist Dammann and named after him [10]. The initial design intent of this grating was to produce multiple beams of diffracted light of equal intensity with extremely uniform distribution among the beams. However, traditional structured light optical elements like the DG often require complex optical systems and high-performance computing resources when applied technologically, posing significant challenges for real-time and portable applications. Moreover, although these conventional optical components can generate structured light, their functionality is relatively fixed and lacks dynamic control, which proves insufficient in applications that require real-time modifications of the light field. Metasurfaces, innovative optical elements whose properties are determined by the microscopic structure of their surfaces arranged at a sub-wavelength scale, can efficiently control the propagation of light waves across them [11]. The optical properties of metasurfaces are dictated by their microstructures, not by the materials that make them up. Metasurfaces exhibit various unique optical properties, such as negative refraction [12], superlensing effects [13], and nonreciprocal transmission [14], which endow them with broad potential in optical applications. By integrating DGs with metasurfaces, it is possible to design a DG metasurface that offers compact structure, high efficiency, and low manufacturing complexity, meeting the demands for miniaturization and integration of structured light generating devices [15-16].

However, a major challenge inherent in metasurfaces is their limited capability for dynamic tuning. Although the miniaturization and integration properties of metasurfaces provide unique advantages in the design of optical components, this limited dynamic tuning capability also restricts their practicality across broader application scenarios. At this juncture, the concept of PT symmetry, a crucial notion in physics, becomes particularly significant. In the field of optics, by introducing PT symmetry, precise control over the gain and loss of light waves can be achieved [17-20]. If we can incorporate PT symmetry into metasurfaces by modulating the loss and gain of nano-pillars through doping and pumping light [21-22], dynamic control of metasurfaces could be realized. This means that by adjusting the state of PT symmetry in the system, we can dynamically alter the structured light generated by the metasurfaces, thus enabling real-time control over three-dimensional imaging. Compared to traditional metasurfaces, the design of PT-symmetric metasurfaces not only allows for nonreciprocal light propagation [23], but also significantly enhances the performance of sensing technologies [24], and exhibits special effects such as anomalous refraction and reflection [25]. These notable characteristics demonstrate irreplaceable value in applications such as optical imaging, optical communications, and sensor technology.

Moiré effect is a phenomenon that produces light-based interference and diffraction patterns when two sets of stripes or dots with differing frequencies, sizes, or spacings are overlaid, resulting in alternating light and dark stripes [29]. By stacking two layers of PT-symmetric metasurfaces, the Moiré effect can be introduced, thereby enhancing the dynamic tuning capability of the entire metasurface. Utilizing the Moiré effect to introduce new optical changes in bilayer metasurfaces, by altering the relative positions and angles between the two layers, enables more advanced dynamic control over structured light. Compared to single-layer tunable metasurfaces, Moiré metasurfaces exhibit significant advantages in continuous wavefront control, expanded tuning range, enhanced performance stability, and simplified operational processes [30-32]. They not only provide more precise wavefront control capabilities but also offer broader modulation ranges and optimized interfaces, making their application in complex optical systems more reliable and convenient.

Thus, in this study, we have designed a novel, dynamically tunable dual-layer DG metasurface by integrating PT symmetry with the Moiré effect. By simply modulating the loss and gain of nano-pillars or rotating one of the layers within the metasurface, we can achieve variations in the pattern of structured light, thereby enabling dynamic adjustability to a certain extent. This design not only facilitates precise control over structured light but also adapts to various application scenarios, boasting wide-ranging practical applications.

## 2. Structure and principle

Firstly, we constructed a model of a double-layer DG consisting of nanocolumns, as shown in Fig. 1(a). In this model, we employed two identical nanocolumn units, symmetrically positioned. These nanocolumn units are based on a silica substrate, while the nanocolumns themselves are made of LN. The base dimension $C$ is $600nm \times 600nm$, the height of the nanocolumns $H$ is $1000nm$, and the distance between the double-layer nanocolumns is $100nm$.

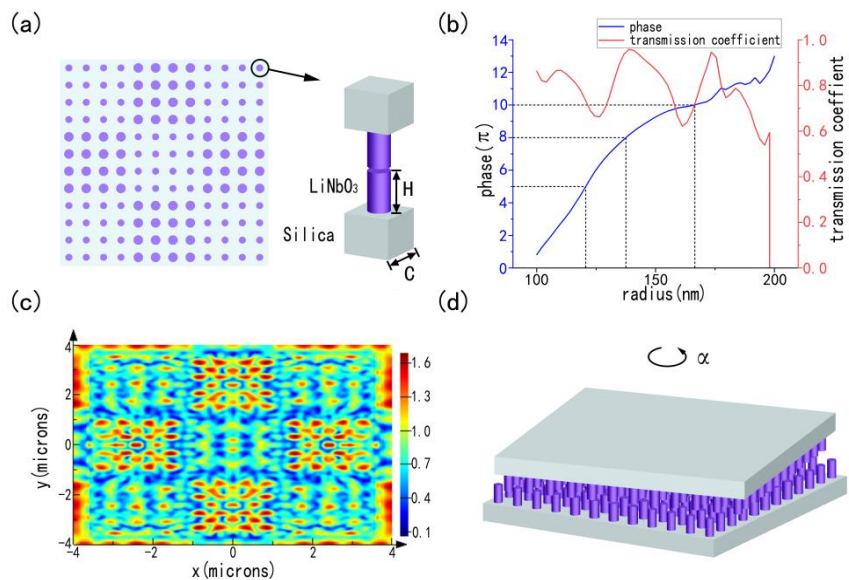

Fig. 1 (a) The top view of the second radius combination double-layer DG metasurface and schematic diagram of the double-layer unit nanocolumn structure. (b) Curves of phase and transmission changes when the radius $R$ varies from $100nm$ to $200nm$. (c) Simulation results of the $E_x$ phase distribution in the transmission field for the second radius combination double-layer DG metasurface. (d) Three-dimensional illustration of the double-layer DG metasurface, where $\alpha$ represents the rotation angle between the two metasurfaces.

To precisely fabricate nanostructures with high aspect ratios and steep sidewalls, we utilized Focused Ion Beam (FIB) milling, a technique particularly suited for producing such structures [30-32]. Nanorods with a length of $100nm$ were selected to balance the technical requirements and design needs. Although the high aspect ratio increases the complexity of the manufacturing process, it also provides us with enhanced phase control capabilities, which are crucial for our design. LN is a material with excellent photonic properties, offering high tunability and significant photorefractive effects, hence its extensive use in optical modulators and other linear and nonlinear optical applications [33-34]. Metasurfaces made from LN can easily achieve a full $2\pi$ phase shift while maintaining high efficiency. Consequently, nanocolumns composed of LN produce varying phase shifts at different radii, thereby altering the phase distribution of the incident light wavefront.

To meet the phase transition requirements of the DG, we performed nested scanning on the double-layer nanocolumn units and illuminated them with a plane wave at a wavelength of $600nm$. Through this method, we obtained phase maps and refractive indices at different radii of the nanocolumns, as depicted in Fig. 1(b). During this process, we selected two sets of radii: $R_1 = 120.632nm$, $R_2 = 137.57nm$ and $R_1 = 120.632nm$, $R_2 = 166.209nm$. For these combinations of radii, the phase differences for the double-layer nanocolumns are $3\pi$ and $5\pi$, respectively, fulfilling the DG's requirement for a $\pi$ phase difference at varying radii.

To generate high-quality structured light, we employed the DG arrangement designed in previous studies [35]. As shown in Fig. 1(a), after arranging the second set of radii, the overall dimensions of the resulting DG metasurface were merely $7.2\mu m \times 7.2\mu m$. Fig. 1(c) displays the electric field $E_x$ transmission phase distribution of the double-layer DG metasurface for the second radius combination under the illumination of a plane wave at $490nm$ wavelength. It is clearly observable that the phase difference of the electric field across different radii is approximately $\pi$, which aligns perfectly with our expectations. For the first radius combination, the $E_x$ transmission phase distribution mirrors that of the second, with phase differences also around $\pi$ across different radii. Fig. 1(d) presents the stereoscopic view of the entire double-layer DG metasurface, illustrating that the layers can be rotated by an angle of $\alpha$, offering our model increased flexibility and adaptability.

At the same time, we have introduced PT symmetry into the DG metasurfaces. Here, we may incorporate loss into one LN nanocolumn by doping with iron, while the other LN nanocolumn gains optical amplification through optical parametric amplification induced by pump light. This arrangement places the entire system in a non-Hermitian state, where gain and loss are manifested in the imaginary parts of the refractive indices of the two nanocolumns. By adjusting the doping concentration of iron and altering the intensity of the pump light, we can equalize the magnitudes of the imaginary parts of the refractive indices of the two nanocolumns while having opposite signs, thereby rendering the real part of the system's refractive index even symmetric and the imaginary part odd symmetric, achieving PT symmetry.

Based on the properties of PT symmetry, we have designed three PT-symmetric states for the double-layer DG Figs. 2(a)–2(c) display the three PT symmetry states implemented in the DGs with the first radius combination, with the second radius combination following the same method for imparting PT symmetry states. Initially, in a single-layer DG, two adjacent nanocolumns of the same radius are treated as one PT symmetry unit, and both layers exhibit identical PT symmetry distributions, as shown in Fig. 2(a). Subsequently, we assign gain or loss separately to the nanocolumns of the upper and lower layers of the DG, such that a nanocolumn pairs with its counterpart in the opposite layer to form a PT-symmetric unit, as illustrated in Fig. 2(b). Finally, based on the first scenario, we modify the PT symmetry of one layer so that a nanocolumn can form a PT-symmetric unit both with an adjacent nanocolumn within the same layer and with a nanocolumn in the other layer, as shown in Fig. 2(c). Referencing the experiment on the complex refractive index of iron-doped LN, we adjusted the imaginary part of the refractive index to 0.1 [36]. The three different double-layer nanocolumn units corresponding to these PT symmetry states are shown in Figs. 2(d)–2(f). We conducted nested scanning on these three types of double-layer nanocolumn units using a plane wave at $800nm$ wavelength, obtaining phase and refractive index maps for the different radii of the three double-layer nanocolumn units. The results demonstrate that these double-layer nanocolumn units maintain an approximate $\pi$ phase difference variation under the two types of radius combinations, despite their differing PT symmetry states.

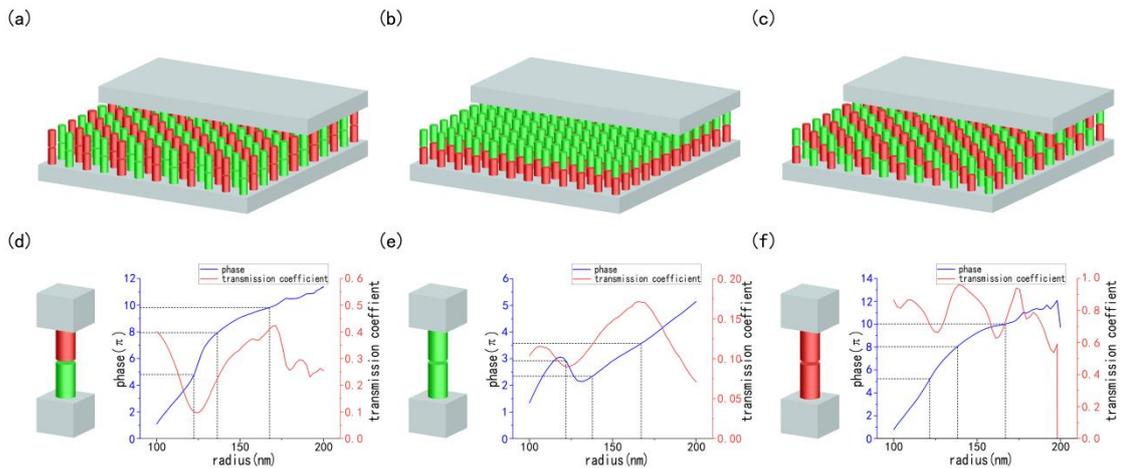

Fig. 2 (a)-(c) Three-dimensional illustrations of the double-layer DG metasurfaces in the first, second, and third PT symmetry states, respectively. (d)-(f) Schematic diagrams of the double-layer nanocolumn unit structures and their phase and transmission change curves as the radius R varies from $100nm$ to $200nm$, corresponding to the first, second, and third PT symmetry states, respectively. The three different double-layer nanocolumn units are shown in the left column of (d)-(f).

The Moiré effect informs us that a rotated grating will exhibit optical properties different from those of the original grating[29]. Illuminating it with the same light source used previously results in diffraction patterns distinct from those produced by an unrotated metasurface. This phenomenon provides us with a novel method to modulate and control the propagation characteristics of light. Taking the DG from the first group of radius combinations in the Hermitian state shown in Fig. 1(d) as an example, we performed rotations of $\alpha = 5°$, $10°$, $15°$, and $20°$ on one layer of the DG metasurface. We observed that these rotations significantly altered the optical properties of the grating. Notably, this rotation effect is present not only in the Hermitian state but also in other PT symmetry states, as well as under the second radius combination, with similar effects observed across these conditions.

## 3. Results and discussion

For the DG metasurfaces with the aforementioned two radius combinations, their performance can be assessed by analyzing and comparing their conversion efficiency and contrast ratio. Referring to previous literature [35], we can directly employ the equations

$$\eta = I_0 + 2\sum_{i=1}^{N} I_i \tag{1}$$

and

$$C = \frac{I_{max} - I_{min}}{I_{max} + I_{min}} \tag{2}$$

to calculate the conversion efficiency and contrast ratio.

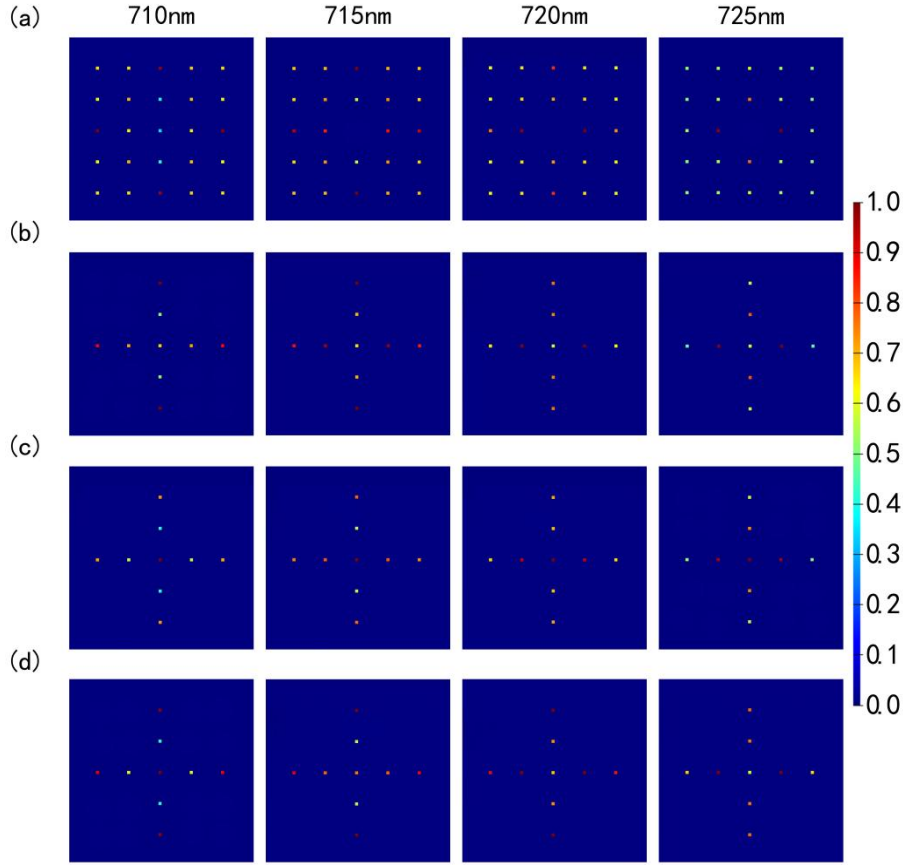

Fig. 3 (a) Normalized intensity distribution of the diffraction spots for the first radius combination double-layer DG metasurface in the Hermitian state under plane wave incidence at $710nm - 725nm$. (b)-(d) Normalized intensity distributions of the diffraction spots for the first radius combination double-layer DG metasurface under plane wave incidence at $710nm - 725nm$, corresponding to the first, second, and third PT symmetry states, respectively.

Initially, we studied the double-layer DG metasurface with the first radius combination. We illuminated it with a plane wave in the $710nm - 725nm$ wavelength range, and the resulting diffraction pattern is shown in Fig. 3(a). It can be observed that this double-layer DG metasurface produces a nearly uniform and equal-intensity $5 \times 5$ diffraction spot array within this spectral band, also demonstrating good conversion efficiency and contrast ratio. Subsequently, we endowed this metasurface with the three described PT symmetry states and modulated the imaginary part of the refractive index to $\pm 0.1$. We again used a plane wave at $710nm - 725nm$, obtaining the near-field diffraction spots, depicted in Figs. 3(b)-3(d). Here, it is noted that we call it the near-field diffraction in contrast to the more distant far-field diffraction in Fig. 7. The results reveal that upon introducing PT symmetry, the light patterns change irrespective of the PT symmetry state considered. The original $5 \times 5$ array transforms into a cross-shaped pattern; however, they differ in the spectral bands that can generate structured light and in the conversion efficiency and contrast ratio of the structured light produced. We define these as the response-sensitive wavelength bands for each state concerning that structure. The comparison of the conversion efficiency and contrast ratio of the diffraction spots produced by the double-layer DG metasurface with the first radius combination in various states is shown in Fig. 4. Experimental data indicate that, after endowing with the three PT symmetry states, the grating continues to exhibit good conversion efficiency and maintains commendable contrast ratio within the range $0.2 - 0.3$ in its response-sensitive bands. These results provide us with crucial information about optimizing metasurface performance through adjusting PT symmetry states and radius combinations. Similarly, the double-layer DG metasurface with the second radius combination, after being endowed with PT symmetry states, showed similar effects. This further validates the efficacy and versatility of our research approach.

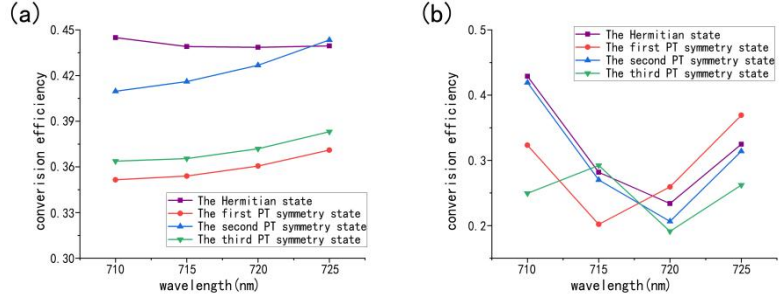

Fig. 4 Comparison of the conversion efficiency (a) and contrast ratio (b) of diffraction spots produced by the double-layer DG metasurface with the first radius combination under different conditions.

To further explore the impact of the Moiré effect on our designed metasurfaces, we conducted rotational transformations on DG metasurfaces with two different radius combinations. This transformation introduces additional phase differences, altering the diffraction pattern and providing a new method of control. For the first radius combination of the DG, in the Hermitian state, we illuminated it with a plane wave at $720nm$. Subsequently, we rotated one layer of the DG metasurface by $\alpha = 5°$、 $10°$、 $15°$、 $20°$. The resulting diffraction spots are displayed in Fig. 5(a), where it is evident that the diffraction pattern changes significantly with increasing rotation angle. Following this, we conducted the same rotational operations on the three PT symmetry states of this metasurface. Using plane waves at wavelengths of $715nm$, $720nm$, and $720nm$ for illumination, the diffraction spots after rotation are shown in Figs. 5(b)-5(d). These results indicate that the diffraction patterns of the metasurfaces under the three PT symmetry states also undergo changes in their light patterns, which are similar across these states yet distinct from changes in the Hermitian state. Under Hermitian conditions, as the angle increases from 5° to 20°, the basic outline of the shapes formed by points with stronger diffraction intensities remains stable, but there are slight changes in the positions and numbers of other diffracted spots. In the PT-symmetric state, with the change in angle, there is a slight variation in the compactness of the overall shape of the diffracted spots, and the positions of the points with stronger diffraction intensities also continuously change. These results further demonstrate that rotational transformations can effectively alter the diffraction patterns of metasurfaces, offering a novel means of manipulation.

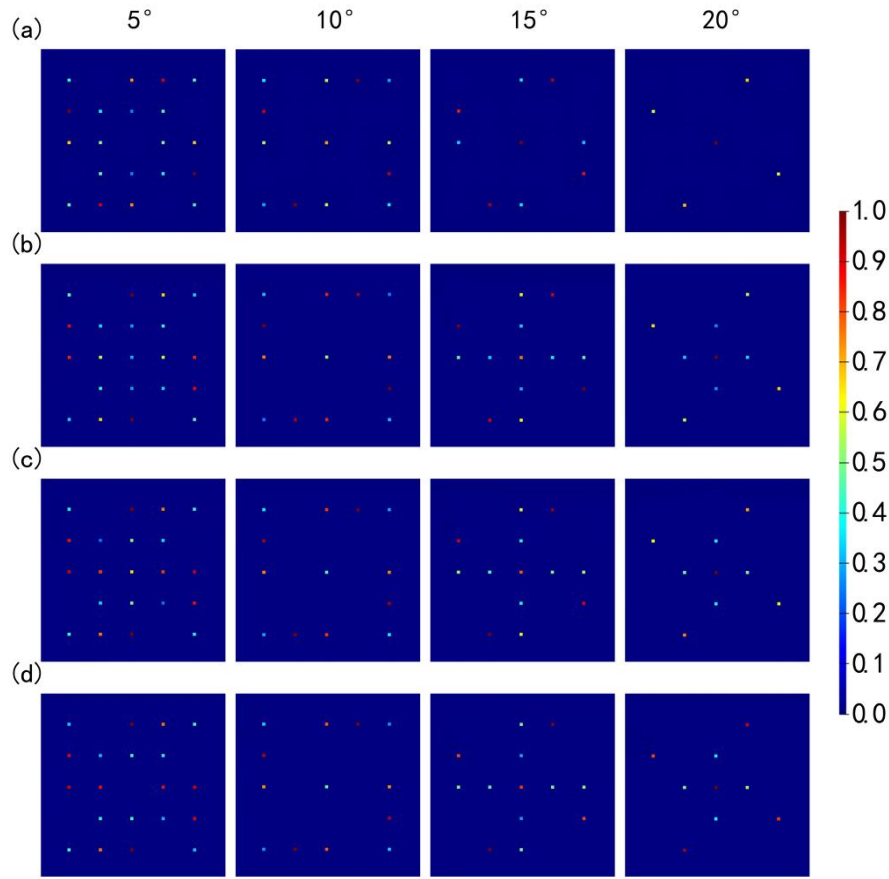

Fig. 5 (a) Normalized intensity distribution of the diffraction spots from the first radius combination DG metasurface in the Hermitian state after rotation. (b)-(d) Normalized intensity distributions of the diffraction spots from the first radius combination DG metasurface after rotation in the first, second, and third PT symmetry states respectively.

Next, we studied the second radius combination of the DG metasurface. In the Hermitian state, we illuminated it with a plane wave at a wavelength of $500nm$. We then rotated a layer of this DG metasurface by $\alpha = 5°$, $10°$, $15°$, and $20°$. The resulting diffraction spots are displayed in Fig. 6(a). Following this, we subjected this metasurface to the same rotational operations across its three PT symmetry states. Illumination was provided by plane waves at wavelengths of $690nm$, $695nm$, and $695nm$ respectively, and the diffraction spots obtained post-rotation are shown in Figs. 6(b)-6(d). Comparing Fig. 5 and Fig. 6, it is clearly observable that both radius combinations of the double-layer DG metasurfaces undergo changes in light patterns after rotation, both in Hermitian and PT symmetry states. The patterns of light change similarly across the PT symmetry states are distinct from those in the Hermitian state, yet vary in shape. This change is evidently due to the effect of Moiré interference. However, the changes in light patterns between the two radius combinations are different. This difference arises because rotating one of the layers of the DG metasurface alters the effective grating spacing, thereby changing the value of the grating period $d$ in the diffraction equation

$$d \sin \theta = m\lambda ,   \quad (3)$$

which in turn modifies the diffraction angle $\theta$. If the diffraction order $m$ and incident wavelength $\lambda$ are unchanged, the different $d$ values for the two radius combinations, due to their differing radii, lead to variations in diffraction angles $\theta$ and consequently in the positions of the light spots, resulting in distinct light patterns. These results provide critical insights into how rotational transformations can be utilized to manipulate the diffraction patterns of metasurfaces.

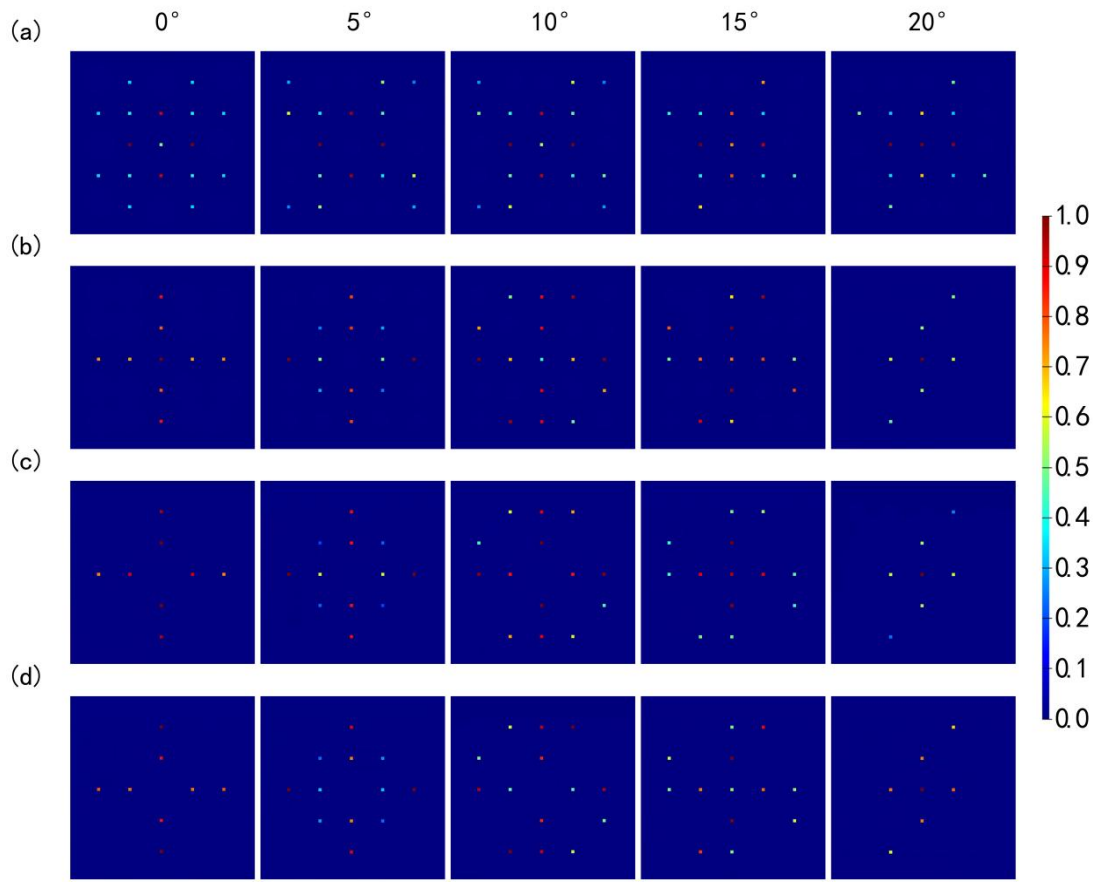

Fig. 6 (a) Normalized intensity distribution of the diffraction spots from the second radius combination DG metasurface in the Hermitian state after rotation. (b)-(d) Normalized intensity distributions of the diffraction spots from the second radius combination DG metasurface after rotation in the first, second, and third PT symmetry states respectively.

To delve deeper into the impact of the Moiré effect on double-layer DGs, we captured far-field diffraction spot views, as shown in Fig. 7. The results indicate that as one layer of the metasurface rotates, there are ongoing changes in the pattern of the light, along with variations in the intensity and positions of different diffraction spots. Specifically, from the figure, we can observe that in the Hermitian state, the double-layer DG metasurface undergoes subtle changes during rotation, with the primary diffraction spots concentrated in the inner ring. In contrast, for the other three PT symmetry double-layer DGs, both the intensity and position of the diffraction spots within the inner and outer rings change. To more clearly illustrate this variation, we select point $M$ on the outer ring of one of the diffraction spot patterns in Fig. 7. Using the center of the diffraction spot as the origin of the coordinate system, we can determine the angles between the two points and the x-axis. Subsequently, by plotting the changing angles of point $M$ with respect to the x-axis under continuous rotational adjustments, as shown in Fig. 8, we observed that the angle variations of point $M$ linearly correlate with the rotation angles. Moreover, we evaluated performance metrics such as conversion efficiency and contrast ratio during rotation, as displayed in Figures 8(b) and 8(c). These findings indicate that although there are some changes in the conversion efficiency across these four states during rotation, the slight variations do not significantly affect the use of structured light. However, an initial decrease in contrast ratio of the diffraction spots is notable at the onset of rotation, with improvements appearing around rotations of $10°$ and $15°$.

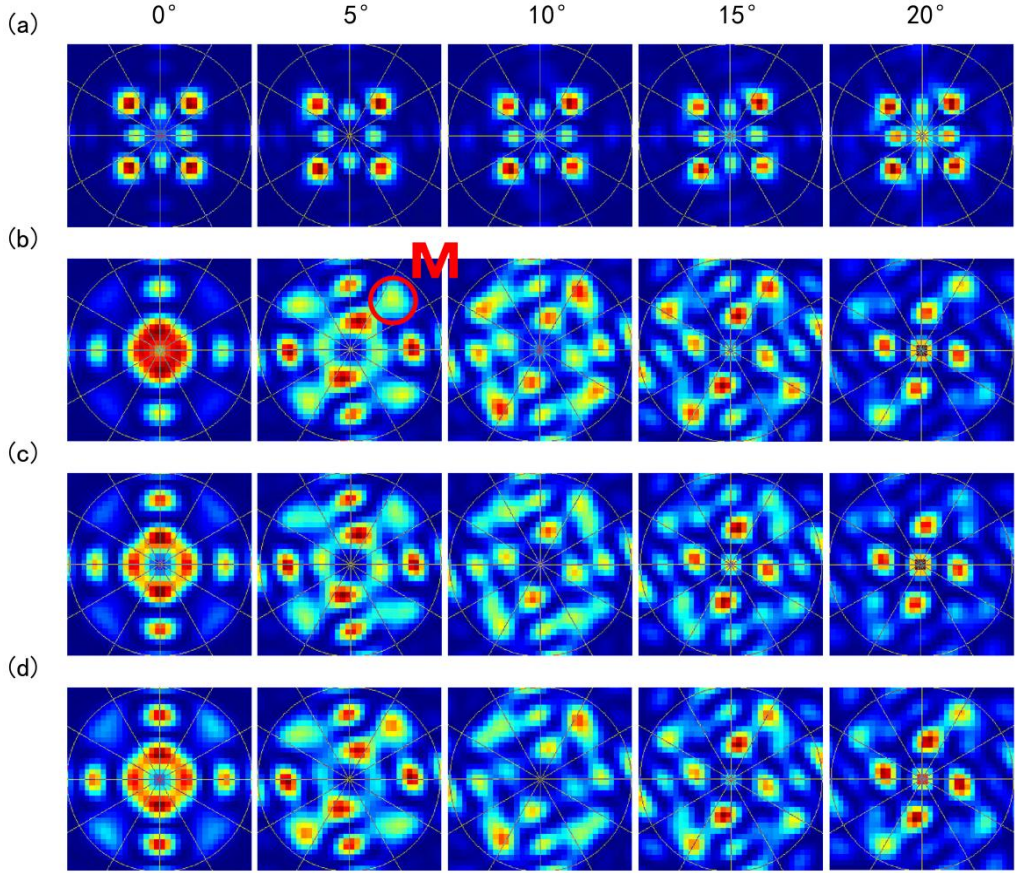

Fig. 7 (a) Far-field spot pattern of the second radius combination DG metasurface in the Hermitian state after rotation. (b)-(d) Far-field spot patterns of the second radius combination DG metasurface after rotation in the first, second, and third PT symmetry states respectively.

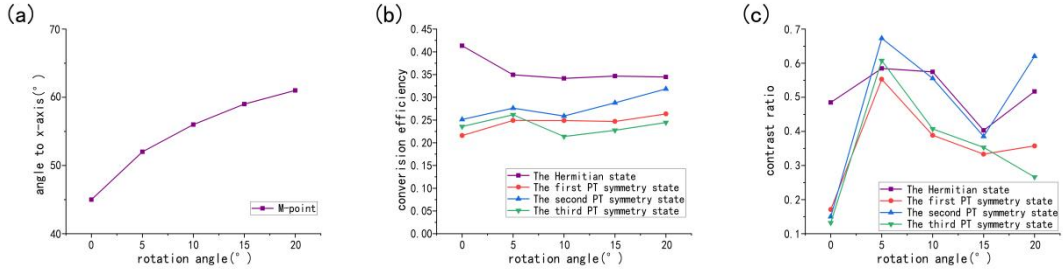

Fig. 8 (a) Angular variation of the angle between point $M$ in the diffraction spot and the x-axis under continuous angular changes. (b) Comparison of conversion efficiency of the second radius combination DG metasurface after rotation in four different states. (c) Comparison of contrast ratio of the second radius combination DG metasurface after rotation in four different states.

## 4. Conclusions

In this study, we comprehensively investigated the optical behavior of double-layer DG metasurfaces influenced by the Moiré effect, with particular focus on their performance under various PT symmetry states.

The findings demonstrated that rotation of one metasurface layer induces changes in the shape and position of diffraction spots, occurring not only in the Hermitian state but also in three PT symmetry states. Initially, for two different radius combinations of double-layer DG metasurfaces, distinct variations in light patterns were observed during rotation. Moreover, although rotation induced slight changes in conversion efficiency, these modifications had a minimal overall impact on the usage of structured light. Notably, when the rotation angle was approximately 10° to 15°, the conversion efficiency and contrast ratio exhibited superior performance, indicating that an optimal selection of rotation angles can enhance the optical properties of the gratings. Additionally, the influence of PT symmetry cannot be overlooked, as it plays a crucial role in modulating the distribution and performance of the diffraction light field. By adjusting the PT symmetry states, further optimization of the light field distribution and performance is achievable. In summary, this research elucidates the profound effects of rotational transformations, PT symmetry, and different radius combinations on the optical characteristics of double-layer DG metasurfaces, providing valuable theoretical insights for future optical applications. These insights not only enrich our understanding of the optical behavior of gratings, but also furnish practical guidance for the design and optimization of grating structures.

## Acknowledgements

Funding: This work was supported by the Natural Science Foundation of China (52175457).

## Author Contributions

Conceptualization, Z. S.; methodology, X. C.; validation, X. C. and Z. S.; formal analysis, X. C., W. L. and Z. S.; investigation, X.C. and H. L.; writing—original draft preparation, X.C.; writing—review and editing, Z. Y. and Z. S.; funding acquisition, H. J. and Y. D. All authors have read and agreed to the published version of the manuscript.

## Data availability

The data presented in this study are available on request from the corresponding author. The data are not publicly available due to privacy restrictions.

## Declarations

### Conflict of interest

The authors declare no competing interests.